\documentclass[12pt]{article}
\usepackage{amsfonts,amssymb,amsmath,eucal,pstricks,pst-plot,pst-node}
\usepackage[dvips]{graphics}

\numberwithin{equation}{section}

\newcommand{\fS}{\mathfrak{S}}
\newcommand{\Z}{\mathbb{Z}}
\newcommand{\R}{\mathbb{R}}
\newcommand{\Q}{\mathbb{Q}}

\newcommand{\C}{\mathbb{C}}
\renewcommand{\P}{\mathbb{P}}
\newcommand{\LV}{{\bigwedge}^{\!\frac{\infty}{2}} V}

\newcommand{\gl}{\mathfrak{gl}}
\newcommand{\fM}{\mathfrak{M}}

\newcommand{\Lc}{\mathcal{L}}
\newcommand{\cE}{\mathcal{E}}

\newcommand{\cS}{\mathcal{S}}
\newcommand{\cZ}{\mathcal{Z}}
\newcommand{\Mc}{\mathcal{M}}
\newcommand{\bK}{\mathsf{K}}
\newcommand{\Mb}{\overline{\Mc}}
\newcommand{\lang}{\left\langle}
\newcommand{\rang}{\right\rangle}
\newcommand{\pt}{\textup{pt}}
\newcommand{\bp}{\mathbf{p}}
\newcommand{\fb}{\mathbf{f}}

\newtheorem{Theorem}{Theorem}

\DeclareMathOperator{\tr}{tr}
\DeclareMathOperator{\Hilb}{Hilb}
\DeclareMathOperator{\Prob}{Prob}
\DeclareMathOperator{\mo}{mod}
\DeclareMathOperator{\grad}{grad}
\DeclareMathOperator{\const}{const}

\begin{document}


\title{The uses of random partitions}

\author{Andrei Okounkov}
\date{July 2003}

\maketitle


\maketitle

\begin{abstract}
 These are extended notes for my talk at the ICMP 2003
in Lisbon. Our goal here is to demonstrate how natural and 
fundamental random partitions are from many different points 
of view. We discuss various natural measures on 
partitions, their correlation functions, limit
shapes, and how they arise in applications, 
in particular, in the 
Gromov-Witten and Seiberg-Witten theory. 
 
\end{abstract}


\section{Recognizing random partitions}

\subsection{Why partitions ?}

\subsubsection{}

Random partitions occur in mathematics and physics in a
wide variety of contexts. For example, a partition can
record a state of some random growth process. More 
often it happens that a certain quantity of interest is
expressed, explicitly or implicitly, as a sum over
partitions. This can come as a result of a localization 
computation in geometry, or from a character expansion 
of a matrix integral, or from something as innocent as
expanding a determinant. Typically, one can recognize in such 
a sum a discrete version of some random 
matrix integral and so one can ask whether 
the powerful and honed tools of the random 
matrix theory can be applied. 

The purpose of these notes is to argue that
certain natural measures on partitions are
not just discrete caricatures of 
random matrix ensembles, but are, in fact,
objects of fundamental importance, with profound connections to 
many central themes of mathematics and 
physics, including, in particular, integrable 
systems. Therefore, I believe that it is very natural
to present these views in this special session on 
\emph{Random Matrix Theory and Integrable Systems}. 

\subsubsection{}

The wealth of applications and connections
of random matrices is such that it is 
utterly impossible to argue that something
is ``just as good'' in one short talk. 
So, instead of trying to paint the whole
picture, I will give a few illustrative
examples, selected according to 
my own limited expertise and taste. Much, much more
can be found in the works cited in 
the bibliography. But even though I tried to 
make the bibliography rather extensive, it is 
still hopelessly far from being complete. 

Several topics that should be covered in any 
reasonable survey on random partitions are
completely omitted here. These include,
for example, the 2-dimensional Yang-Mills
theory and character expansions in the 
random  matrix theory. We also say nothing
about the relation between random partitions
and planar dimer models, even though the 2-dimensional
point of view on random partitions if often 
illuminating, has several technical advantages, as
well as some exciting connections to algebraic geometry
\cite{KOS}. 

\subsection{Coordinates on partitions}

\subsubsection{Diagram of a partition}

By definition, a partition $\lambda$ is simply a 
monotone sequence 
$$
\lambda=(\lambda_1\ge \lambda_2 \ge \lambda_3 \ge \dots \ge 0)
$$
of nonnegative integers such that $\lambda_i=0$ for $i\gg 0$.
The \emph{size} of $\lambda$ is, by definition,
the number $|\lambda|=\sum \lambda_i$.  
 
\begin{figure}[!htbp]
  \begin{center}
    \scalebox{0.8}{\includegraphics{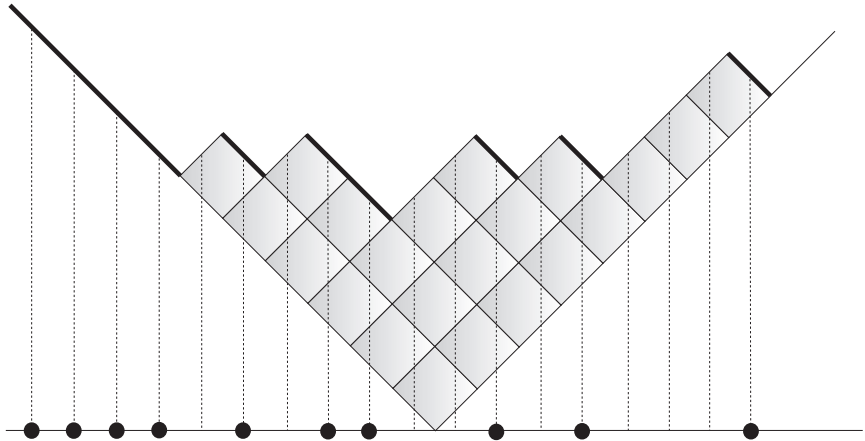}} 
    \caption{The partition $(8,5,4,2,2,1)$}
    \label{f1}
  \end{center}
\end{figure}

The standard geometric object associated to a partition 
is its \emph{diagram}. There are several competing traditions
of drawing the diagram. We will follow the one illustrated in 
Fig.~\ref{f1}, 
which  portrays the partition
$\lambda=(8,5,4,2,2,1)$. 
This way of drawing diagrams is sometimes referred to as the Russian
one (as opposed to the older French and English traditions of
drawing partitions). Its advantages are
not just that the picture looks more balanced on 
the page and saves space by a factor of $\approx\sqrt{2}$,
but also that from it one can see more clearly several 
other useful ways of parameterizing the partitions.

\subsubsection{Profile of a partition}

The upper boundary in Fig.~\ref{f1} is a
graph of a function $f_\lambda(x)$
such that $f'_\lambda(x)=\pm 1$. This function $f_\lambda$ is 
known as the \emph{profile} of the partition $\lambda$. 
The map $\lambda\mapsto f_\lambda$ from partitions to 
functions with Lipschitz constant $1$ allows one to talk 
about \emph{limit shape} of partitions. Namely, given a
sequence of probability measures on partitions, we say
that it has a limit shape $f$ if, after a suitable scaling,
the corresponding measures on functions 
converge weakly to the $\delta$-measure on $f$.

\subsubsection{Partitions vs.\ particles}

Another useful way to parametrize partitions is 
via the map 
$$
\fS: \lambda \mapsto \left\{\lambda_i-i+\frac12\right\} \subset \Z+\frac12
$$
from partitions to subsets of $\Z+\frac12$. The geometric 
meaning of the set $\fS(\lambda)$ should be  clear 
from Fig.~\ref{f1} where the elements of $\fS(\lambda)$ 
are shown by bold dots. The map $\fS$ makes a random
partition a random subset of $\Z+\frac12$ or, in other
words, an \emph{ensemble of random particles}  on the lattice. 
It is these random particles that are the 
analogs of eigenvalues of a random matrix. A natural
question here is to compute the \emph{correlation functions}, that 
is, the probability to observe particles in specified locations.

\subsection{Partitions and the fermionic Fock space}

\subsubsection{Subspaces instead of subsets}

Note that, for any partition $\lambda$, the set $\fS(\lambda)$
has as many positive elements as it has negative holes
(that is, negative half-integers not in $\fS(\lambda)$). 
It can be viewed, therefore, as a finite excitation 
over the Dirac sea
$$
\fS(\emptyset) = \left\{-1/2,-3/2,-5/2,\dots \right\} \,,
$$
in which all negative positions are filled by particles
while all positive ones are vacant. 

Consider the vector space $V$
with an orthonormal  basis $\{e_k\}$, $k\in \Z+\frac12$, indexed by 
all possible positions of one particle. To a partition
$\lambda$ one then associated the following vector 
$$
v_\lambda = e_{\lambda_1-\frac12} \wedge 
e_{\lambda_2-\frac32} \wedge 
e_{\lambda_3-\frac52} \wedge 
\dots 
$$
in the half-infinite exterior power $\LV$
of the $1$-particle space $V$. In other words, one associates 
to a partition $\lambda$ 
 the image of the subspace spanned by $\{e_k\}$, $k\in \fS(\lambda)$,
under the Pl\"ucker embedding of the corresponding Grassmannian.

Note that the vectors
$v_\lambda$ are orthonormal with respect to 
the natural inner product, and, in particular, any
vector $v$ in their span defines a probability measure
measure $\fM_v$ on partitions by 
\begin{equation}
  \label{Mv}
  \fM_v(\lambda) = \frac{|(v,v_\lambda)|^2}{\|v\|^2} \,.
\end{equation}

\subsubsection{The action of $GL(\infty)$}

The main advantage of trading sets for linear
spaces like we just did is the following. 
The group $GL(V)$ of invertible linear
transformations of a finite-dimensional 
vector space $V$ acts naturally in all
exterior powers of $V$. The case of an 
infinite exterior power of an infinite-dimensional
space requires more care (in particular,
the need for normal ordering of operators
arises), but it is still possible to  define a 
projective action on $\LV$ of a suitable version  
of the group $GL(V)$, see for example \cite{K,MJD,SV}. 

For our purposes, it suffices to define
the action of the operators of the form $e^X$, where the matrix
$X$
lies in the Lie algebra $\gl(V)$. If
$X$ has zeros on the diagonal then the 
naive definition of its action on $\LV$ works
fine (note however, that a central extension appears,
see e.g.\ \eqref{commal}). For diagonal matrices $X$, we will use the
formula \eqref{pkE} as our regularization 
recipe.

It is 
the gigantic symmetry group $GL(V)$ that makes
certain computations in $\LV$ so pleasant 
(such as, for example, computations of
correlation functions, see below). It 
also opens up the connection to 
integrable systems, where, through the
work of the Kyoto school and others, the space $\LV$
has become one of the cornerstones of the
theory. 


\subsection{Plancherel measure}

We conclude this introductory section with 
the discussion of the most basic measure
on the set of partitions --- the Plancherel 
measure.

\subsubsection{Representation theory}

The general linear group and the symmetric
group $S(n)$ are the two most important 
groups in mathematics and the representation 
theory of both groups is saturated with partitions. 
In particular, over a field of characteristic zero (we 
will not come across any other fields in these notes), irreducible
representations of $S(n)$ are indexed by partitions
$\lambda$ of size $n$. Let $\dim \lambda$ denote 
the dimension of the corresponding representation. 
It follows from a theorem of Burnside
that the formula
\begin{equation}
  \label{defPl}
  \fM_\textup{Planch}(\lambda) = \frac{(\dim\lambda)^2}{n!}
\end{equation}
defines a probability measure on the set of partitions 
of $n$. This measure is known as the \emph{Plancherel 
measure}, because of its relation to the Fourier
transform on the group $S(n)$. 

Often, it is more convenient to consider a
related measure on the set of all partitions
defined by
\begin{equation}
  \label{defPP}
     \fM_\textup{PP}(\lambda) = e^{-\xi}\, \xi^{|\lambda|}\,
\left(\frac{\dim\lambda}{|\lambda|!}\right)^2\,, \quad 
\xi > 0 \,.
\end{equation}

It is known as the \emph{poissonized} Plancherel measure, the number
 $\xi>0$ being the parameter of the poissonization.

\subsubsection{Plancherel and GUE}

While representation theory provides an 
important motivation for the study of the 
Plancherel measure,  the representation--theoretic
definition of it may sound like something very
distant  from the ``real world''
until one recognizes, through other 
interpretations of the number $\dim \lambda$,
that one is dealing here with a distinguished
discretization of the GUE ensemble from the 
random matrix theory. S.~Kerov (see, for example
his book \cite{Keb}) and K.~Johansson \cite{J1}
were among the first to recognize this connection.

A pedestrian way to see the connection to the GUE
ensemble is to use the following formula 
\begin{equation}
  \label{diml1}
\dim \lambda = \frac{|\lambda|!}
{\prod (\lambda_i+k-i)!} \, 
\prod_{i<j\le k} (\lambda_i-\lambda_j+j-i)\,,
\end{equation}
where $k$ is any number such that $\lambda_{k+1}=0$. 
The first factor in \eqref{diml1} 
is roughly a multinomial coefficient and
hence a discrete analog of the Gaussian weight,
while the second factor looks like Vandermonde 
determinant in the variables $\fS(\lambda)$. This 
makes $(\dim\lambda)^2$ resemble the radial part
of the GUE measure given, up to a constant
factor,  by the weight
\begin{equation}
  \label{wGUE}
    e^{-\frac12 \sum x_i^2} \prod_{i<j\le N} (x_i-x_j)^2  \,,
\end{equation}
the particles $\fS(\lambda)$
playing the role of the eigenvalues $\{x_i\}$.

\subsubsection{Plancherel measure and random growth}

\begin{figure}[!htbp]
  \begin{center}
    \scalebox{0.6}{\includegraphics{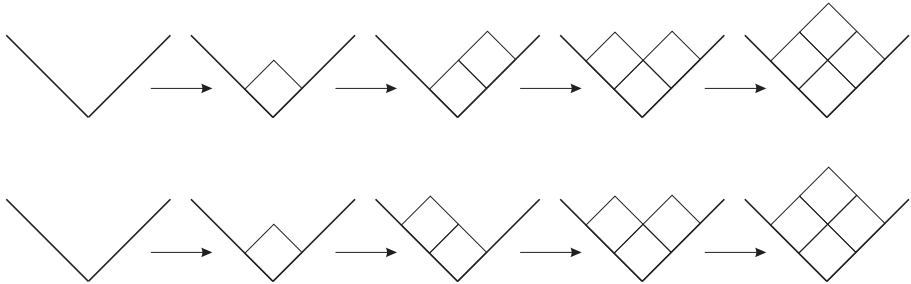}} 
    \caption{The two ways to grow the partition $(2,2)$}
    \label{f2}
  \end{center}
\end{figure}

Another interpretation of $\dim\lambda$ is the following:
it is the number of ways to grow the diagram of $\lambda$ from the 
empty diagram $\emptyset$ by adding a square at a 
time, while maintaining a partition at every step. For 
example, $\dim (2,2) = 2$ corresponds to the two 
possible growth histories shown in Fig.~\ref{f2}. 
In representation theory, this interpretation of 
$\dim\lambda$ is a consequence of  the 
branching rule for the restriction $S(n)\downarrow S(n-1)$. 
It links Plancherel measure
with the Robinson-Schensted algorithm (see, for example, \cite{BF,S})
and many related growth processes. 

In particular, by a 
theorem of Schensted, the distribution of $\lambda_1$ 
with respect to $\fM_\textup{Planch}$ is precisely
the distribution of the longest increasing subsequence
in a uniformly random permutation of $\{1,\dots,n\}$. 
The understanding of this distribution was a major stimulus
for the study of Plancherel measure, culminating
in the work of J.~Baik, P.~Deift, and K.~Johansson \cite{BDJ}. 
They proved that the scaled and centered distribution of $\lambda_1$
converges to the Tracy-Widom distribution \cite{TW} which describes 
the maximal eigenvalue of a large random Hermitian matrix. They 
also proved a similar statement for $\lambda_2$ and 
conjectured that, more generally, the joint 
distribution of $\{\lambda_1,\lambda_2,\dots\}$, scaled and 
centered, converges to the \emph{Airy ensemble} which describes 
the behavior of the $1$st, $2$nd, and so on eigenvalues of 
a large random matrix. This conjecture was established in 
\cite{Orp,BOO,J2}, see more in Section \ref{sBDJ} below. 

Plancherel measure also arises in more general growth 
processes, where both adding and removing a square is
allowed, see below. This is a discrete analog of how one finds the GUE
distribution in the context of non-intersecting
Brownian motions, see for example \cite{J3}.

\subsubsection{Operator form of partition growth}

Consider the following elements of the Lie algebra $\gl(V)$
\begin{equation}
  \label{defal}
    \alpha_n \cdot e_k = e_{k-n} \,. 
\end{equation}
{}From definitions, one finds that in the basis $v_\lambda$ 
the operator $\alpha_{-1}$ acts as follows
\begin{equation}
  \label{aclal1}
  \alpha_{-1} \cdot v_\lambda  = \sum_{\mu=\lambda+\square} v_\mu \,,
\end{equation}
where the summation is over all possibilities to add a square 
to the partition $\lambda$. Exponentiating \eqref{aclal1} and comparing
it to \eqref{Mv}, we 
conclude that $\fM_{PP} = \fM_{v}$, where 
\begin{equation}\label{vPP}
 v = 
\exp\left(\sqrt{\xi} \, \alpha_{-1}\right) \cdot  v_\emptyset \,.
\end{equation}
Similarly, the adjoint operator $\alpha_1=\alpha_{-1}^*$
removes a square from partitions. From the basic commutation 
relation 
\begin{equation}
  \label{commal}
  \left[\alpha_n,\alpha_m\right] = n \, \delta_{n+m} 
\end{equation}
satisfied by the operators $\alpha_n$ in the (projective)
representation $\LV$, one sees that mixing adding squares
with removing squares leads again to the Plancherel measure. 

The formula \eqref{vPP} leads to a simple computation 
of the correlation function of the Plancherel measure,
see below.

\section{Random partitions in Gromov-Witten theory}

\subsection{Random matrices and moduli of curves}

\subsubsection{Matrix integrals in 2D quantum gravity}\label{s2dqg}

\begin{figure}[!htbp]
  \begin{center}
    \scalebox{0.6}{\includegraphics{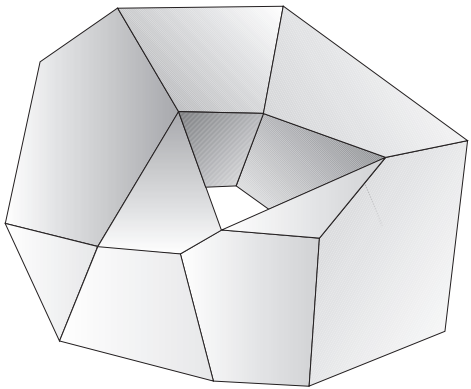}} 
    \caption{A cubist's view of a discretized torus}
    \label{fcub}
  \end{center}
\end{figure}

The Wick formula expansion of the following Gaussian
integral over the space of $N\times N$ Hermitian
matrices $H$
\begin{equation}
  \label{GUWick}
  \int e^{-\frac12 \tr H^2} \prod_{i=1}^n \tr H^{k_i} \, dH  = 
\textup{const} \int_{\R^N} e^{-\frac12 \sum x_i^2}
 \prod_{i<j}
(x_i-x_j)^2 \prod_{i=1}^n \left(\sum_{j=1}^N x_j^{k_i}\right) \, dx 
\end{equation}
is well-known to enumerate different ways to glue 
an orientable surfaces of a given genus $g$ from 
a $k_1$-gon, $k_2$-gon, etc., see for example \cite{Z} for
an elementary introduction. 
In other words, the integral \eqref{GUWick} can be 
written as a certain sum over discretized surfaces,
also known as \emph{maps on surfaces}, 
of the kind shown in Fig.~\ref{fcub}. 
This sum over surfaces can, in turn, be viewed as a Riemann 
sum for a certain vaguely defined ``integral''
 over the infinite-dimensional
space of all metrics on a genus $g$ surface.
In 2-dimensional quantum gravity, one 
wants to integrate over this space of metrics;
this is the reason why integrals \eqref{GUWick} are 
studied there, see, in particular, \cite{BK,Dou,DS,GM1,GM2} and 
\cite{dF,DFGZ} for a survey.

More precisely, the relevant $N\to\infty$ asymptotic regime is 
when the number $n$ of pieces forming the 
surface grows, while the shapes $k_i$ of individual pieces remain 
bounded. For example, one can look at surfaces
composed out of a large number of squares and 
hexagons. The corresponding
(hard) mathematical problem is the $N\to\infty$ 
asymptotics of the integral
\begin{equation}
  \int e^{-\frac12\tr H^2 + \tr P(H)}\, dH \,,
\end{equation}
where $P(H)$ is a polynomial of $H$ the 
coefficients of which depend on $N$ in a
certain critical fashion.

\subsubsection{Witten's conjecture}\label{sWitten}

It was conjectured by Witten \cite{W} that this
discretized ``integration'' over the space of 
metrics is equivalent to certain topologically
defined integrals over the (finite-dimensional)
space of just the conformal classes of smooth
metrics. Conformal, or, equivalently, complex structures on a 
genus $g$ surface form a finite-dimensional
moduli space $\Mc_g$ of dimension $3g-3$ for
$g>1$. The integrals in Witten's conjecture 
are integrals of Chern classes of certain natural line 
bundles and, as a result, they have a topological interpretation as 
intersection numbers on a certain distinguished
compactification $\Mb_g\supset \Mc_g$. 

This compactification $\Mb_g$, constructed by Deligne and Mumford,  
is obtained by
allowing nodal degenerations (of the kind seen on the 
left in Fig.~\ref{fmod}) satisfying certain 
stability conditions, see, for example, \cite{HM}. 
Similarly, the moduli space 
$\Mc_{g,n}$ of smooth genus $g$ algebraic curves $C$
together with a choice of $n$-distinct marked 
points $p_1,\dots,p_n\in C$ has a stable compactification
$\Mb_{g,n}$, which is a projective algebraic variety
of dimension $3g-3+n$. For our purposes, it is 
somewhat more convenient to allow \emph{disconnected} curves $C$.
The disconnected and connected theories are, of course, equivalent. 

By construction, a point of $\Mb_{g,n}$ corresponds to
an at worst nodal curve $C$ with a choice of $n$ smooth 
marked points $p_i\in C$. Since $p_i$ is a smooth point 
of $C$, one can consider the cotangent line $T^*_{p_i} C$
to $C$ at $p_i$. As the reference point in  $\Mb_{g,n}$
varies, these lines form a line bundle $\Lc_i$ over $\Mb_{g,n}$. 
Consider the first Chern class of this bundle
$$
c_1(\Lc_i) \in  H^2(\Mb_{g,n}) \,.
$$
Witten's conjecture (first proved in \cite{Kon})
was that the natural generation function for
the intersection numbers 
\begin{equation}
  \label{deftau}
\lang \tau_{k_1} \dots \tau_{k_n} \rang 
= \int_{\Mb_{g,n}} 
 c_1(\Lc_1)^{k_1} \dots c_1(\Lc_n)^{k_n} \,, \quad \sum k_i= 3g-3+n\,, 
\end{equation}
is precisely the $\tau$-function of the KdV hierarchy 
that emerged from the study of the matrix model 
of the 2D quantum gravity.

\subsubsection{Edge of the spectrum and moduli of curves}

The same $\tau$-function also arises in   
a mathematically much simpler asymptotic regime 
of the integral \eqref{GUWick}, namely the 
one in which the $k_i$'s go to infinity 
simultaneously with $N\to\infty$, while their 
number $n$ remains bounded. It is clear that 
the largest eigenvalues of $H$ dominate 
this asymptotics. Since the distribution of eigenvalues of 
$H$ near the edge of its spectrum is well known 
to converge to the Airy ensemble, this limit
is very easy to control. 

Somewhat surprisingly, it is 
this simpler matrix model that has a 
natural geometric relation to $\Mb_{g,n}$,
as discussed in detail in \cite{OP1} and also, e.g., in  \cite{O2}, leading,
in particular, to another, and in several 
respects more satisfying,  proof of Witten's conjecture.

\subsection{Gromov-Witten theory of $\P^1$}

One explanation of why the 
two very different asymptotic regimes lead
to the same result is the following. 
The intersection theory of $\Mb_{g,n}$
embeds, in fact in many different ways, into a 
certain richer
geometric theory which, as it turns out, 
can be described by the Plancherel 
measure analog of \eqref{GUWick} on a finite
level, that is, \emph{without taking any limits}. 

\subsubsection{Stable maps and GW invariants}

\begin{figure}[!htbp]
  \begin{center}
    \scalebox{0.8}{\includegraphics{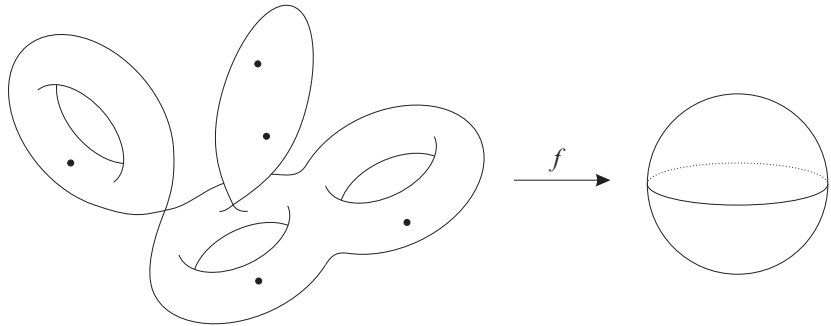}} 
    \caption{A schematic view of a boundary element in $\Mb_{3,5}(\P^1)$}
    \label{fmod}
  \end{center}
\end{figure}

This richer theory is the Gromov-Witten 
theory of the Riemann sphere $\P^1$, which 
is defined using intersection theory on 
the moduli space $\Mb_{g,n}(\P^1,d)$ of
\emph{stable maps} to $\P^1$, see e.g.\ \cite{FuP}.
 By definition, a point in 
the space of stable maps is described by the data
\begin{equation}
  \label{ptstM}
  f: (C,p_1,\dots,p_n) \to \P^1\,,
\end{equation}
where $f$ is a degree $d$ holomorphic map whose 
domain $C$ is a possibly nodal curve of 
genus $g$ with smooth marked points $p_i\in C$, see
Figure \ref{fmod}. 
 Here, again, possible degenerations of the domain $C$ 
are limited by a certain stability 
condition. 

An open (but not dense) subset 
$$
\Mc_{g,n}(\P^1,d)\subset \Mb_{g,n}(\P^1,d)
$$
is formed by maps with smooth
domains $C$.  In this case $f$ represents $C$ as 
a Riemann surface of an algebraic function of
degree $d$. Generically, $f$ has only nondegenerate
critical points, the number of which
equals $2d+2g-2$ by Riemann-Hurwitz. 
The corresponding
critical values together with the $n$ images $f(p_i)\in\P^1$  
of the marked points 
give convenient local coordinates on a 
neighborhood of such map $f$.  
The number $2d+2g-2+n$ is known as the (complex)
\emph{expected
 dimension} of $\Mb_{g,n}(\P^1,d)$.

The whole $\Mb_{g,n}(\P^1,d)$ is not so nice, being
 reducible
with components of different dimensions. One defines,
however, a distinguished homology class
\begin{equation}
  \label{vircla}
   \left[\Mb_{g,n}(\P^1,d)\right]^\textup{vir} \in 
H_*(\Mb_{g,n}(\P^1,d))
\end{equation}
of the expected
dimension, known as the \emph{virtual fundamental class}. 
Integration against \eqref{vircla} replaces in Gromov-Witten
theory integration over fundamental class in \eqref{deftau}. 

The most fundamental part of the Gromov-Witten theory of $\P^1$ is 
its \emph{stationary sector}, obtained by pinning down the images 
of the marked points by requiring $f(p_i)=q_i$,
where $q_i\in \P^1$ are arbitrary distinct points. 
The stationary GW invariants of $\P^1$ are, by definition, the following 
numbers 
\begin{equation}
  \label{deftaupt}
  \lang \tau_{k_1}(\pt)\, \dots \, \tau_{k_n}(\pt) \rang_d 
= \int_{ \left[\Mb_{g,n}(\P^1,d)\right]^\textup{vir} 
\bigcap \{f(p_i)=q_i\}_{i=1\dots n}} 
 c_1(\Lc_1)^{k_1} \dots c_1(\Lc_n)^{k_n} \,, 
\end{equation}
where the classes $c_1(\Lc_i)$ are defined as before. Note that
the genus $g$ is uniquely determined by the dimension
constraint
$$
\sum k_i = 2d + 2g - 2
$$
and, therefore, is omitted in the left-hand side of \eqref{deftaupt}.

\subsubsection{Plancherel measure and GW invariants of $\P^1$}

In order to write down the Plancherel measure analog of the 
matrix integral \eqref{GUWick} we need the partition 
analog of the function 
\begin{equation}
  \label{trace}
  \tr H^k = \sum_{i=1}^N x_i^k \,.
\end{equation}
It is given by the following function 
\begin{align}
  \label{defbp}
   \bp_k(\lambda) 
   &= \sum_i \left[\left(\lambda_i-i+\tfrac12\right)^k -
\left(-i+\tfrac12\right)^k\right] + \left(1-2^{-k}\right) \, \zeta(-k)\\
&\textup{\!``=''} \sum_{x\in \fS(\lambda)} x^k \,, \notag 
\end{align}
where the first line the $\zeta$-regularization of the 
direct, but divergent, generalization of \eqref{trace}
written in second line. Equivalently,
\begin{equation}
   \frac1z + \sum_k \bp_k(\lambda) \, z^k
   =  \sum_i e^{z(\lambda_i-i+\frac12)} 
   = (\cE(z)\cdot v_\lambda, v_\lambda)
  \label{pkE} \,, 
\end{equation}
where $\cE(z)$ is the following diagonal matrix in $\gl(V)$
\begin{equation}
  \label{defcE}
  \cE(z) \cdot e_k = e^{zk} \, e_k \,.
\end{equation}
We now ready to state the following 
result from \cite{OP2}

\begin{Theorem}\label{t1}
  \begin{equation}
    \label{thm1}
     \lang \tau_{k_1}(\pt)\, \dots \, \tau_{k_n}(\pt) \rang_d 
= \frac1{\prod (k_i+1)!}\, \sum_{|\lambda|=d}   
\left(\frac{\dim\lambda}{d!}\right)^2 \, \prod \bp_{k_i+1}(\lambda) \,.
  \end{equation}
\end{Theorem}

As explained above, the right-hand side of \eqref{thm1} is 
a direct analog of the integral \eqref{GUWick} for partitions
of \emph{finite} size $d$, where $d$ is the degree of GW 
invariant on the left. 
In particular, this formula is highly nontrivial already for 
$d=0$, reproducing a result of \cite{FP}. There is only 
one partition of $0$, namely the empty partition $\emptyset$
and $\bp_k(\emptyset)$ equals the $\zeta$-term in \eqref{defbp},
illustrating the naturality of the $\zeta$-regularization. 

Theorem \ref{t1} continues to hold for stationary 
GW invariants of any smooth curve $X$, with the 
following modification:
$$
 \left(\frac{\dim\lambda}{d!}\right)^2 \mapsto \, 
\left(\frac{\dim\lambda}{d!}\right)^{2-2g(X)} \,,
$$
where $g(X)$ is the genus of $X$. 
In particular, genus 1 targets lead to the uniform 
measure on partitions.

\subsubsection{Operator form of the GW theory}

We reproduced here the formula \eqref{thm1} to 
emphasize the role of the Plancherel measure in 
the GW theory. It is, however, very much not
the final answer in the theory. 
Using \eqref{pkE} and \eqref{vPP}, one rewrites
\eqref{thm1} as follows
\begin{equation}
  \label{operthm1}
  \sum_{k_i} \lang \prod \tau_{k_i}(\pt)\rang_d \prod z_i^{k_i+1}
= \frac{1}{(d!)^2} 
\lang \alpha_1^d\,  \prod \cE(z_i)\,  \alpha_{-1}^d 
\rang \,,
\end{equation}
where the angle brackets on the right denote 
the vacuum expectation 
$$
\lang M \rang = (M v_\emptyset, v_\emptyset)\,,
$$
of an operator $M$ acting on $\LV$. 

 Using the commutation relations 
between the operators $\alpha_n$ and $\cE(z)$, one
evaluates \eqref{operthm1} in closed form in terms
of trigonometric functions \cite{OP2}. For example, the 
$1$-point function has an especially simple form 
\begin{equation}
  \label{1ptP1}
  \sum_{g\ge 0} 
\lang \tau_{2g-2+2d}(\pt) \rang^\circ_d \, z^{2g} =
\frac1{(d!)^2} \, \cS(z)^{2d-1} \,, \quad 
\cS(z) = \frac{\sinh z/2}{z/2} \,,
\end{equation}
where the superscript ${}^\circ$ denotes the 
connected GW invariant. 

In the case when 
the target $X$ is an elliptic curve $E$, the 
vacuum matrix element is replaced by trace
\begin{equation}
  \label{gen1tr}
 \sum_{d\ge 0} q^d
 \sum_{k_i} \lang \prod \tau_{k_i}(\pt)\rang_d^E \prod z_i^{k_i+1}
= \tr_0 q^{L_0} \prod \cE(z_i) \,,
\end{equation}
where $L_0$ is the energy operator defined by 
$$
L_0 \cdot v_\lambda = 
|\lambda|\, v_\lambda\,,
$$
and $\tr_0$ denotes the trace in the zero charge subspace
(spanned by the vectors $v_\lambda$). The sum \eqref{gen1tr}
was computed in \cite{BO} in terms of genus 1 theta functions
with modular parameter $q$, see also \cite{Oiw}. 

{}From the operator interpretation, one  
derives Toda equations. These equations
were conjectured in \cite{EgHY,EgY}, see also 
\cite{G1,G2,Pt}, and translate into 
effective recurrence relations for the GW 
invariants of $\P^1$. The integrable structure of the GW
theory fully unfolds in the \emph{equivariant}
GW theory of $\P^1$, which is described 
by the 2-dimensional Toda hierarchy 
of Ueno and Takasaki \cite{UT}. The 2D Toda 
hierarchy is derived from the operator solution 
of the theory in \cite{OP3}. 
The description of the nonstationary sector of
the theory is completed in \cite{OP4}.

\section{More random partitions from geometry}

\subsection{Hurwitz theory}

The Gromov-Witten theory of target curves is closely
related to the much older and much more 
elementary Hurwitz theory \cite{Hur} that concerns
enumeration of degree $d$ branched covers
$$
f: C\to X
$$
of a smooth curve $X$ with specified
ramifications. What this means is: we require $f$
to be unramified outside some fixed set of points
$\{q_i\}\in X$ and for each point $q_i$
we specify the conjugacy class  in the symmetric group $S(d)$
of the monodromy
of the branched cover $f$ 
around $q_i$. In other 
words, for every $q_i$ we specify a partition 
$\eta^{(i)}$ of the number $d$.

By cutting the base $X\setminus\{q_i\}$ (and hence the cover $C$) 
into simply-connected pieces, one can see
many connections between the Hurwitz theory and 
enumeration of maps on $C$ discussed in Section \ref{s2dqg}.

A classical formula of Burnside \cite{Burn,J} tells us that the
number of such covers, automorphism-weighted and  possibly disconnected,
 equals 
\begin{equation}
  \label{Hur}
  \sum_{|\lambda|=d} \left(\frac{\dim\lambda}{d!}\right)^{2-2g(X)} \, 
\prod \fb_{\eta^{(i)}}(\lambda) \,,
\end{equation}
where $g(X)$ is the genus of $X$ and $\fb_\eta(\lambda)$ is the 
\emph{central character} of the representation $\lambda$, that is, 
the unique eigenvalue of the matrix by which the conjugacy 
class $\eta$ acts in the irreducible representation $\lambda$. 

By a theorem of Kerov and Olshanski \cite{KO}
\begin{equation}
  \label{Ls}
  \fb_\eta \in \Lambda^*=\Q[\bp_1,\bp_2,\bp_3, \dots]\,,
\end{equation}
which shows that from the random partitions point of view
there is no real difference between \eqref{thm1} and \eqref{Hur}. 
This is a manifestation of the \emph{Gromov-Witten/Hurwitz
correspondence}, established in \cite{OP2}.

\subsection{Uniform measure and ergodic theory}

Note from \eqref{Hur} that enumeration of 
degree $d$ branched covering of the torus is related to 
the \emph{uniform} measure on partitions of $d$. 
The large $d$ asymptotics in this problem is 
interesting, because, on the geometric side,
it computes the volumes of moduli spaces
of pairs $(C,\omega)$ where 
$C$ is a smooth curve and $\omega$ 
a holomorphic differential on $C$ with 
given multiplicities of zeros \cite{EO}. This is
because points of the form $(C,f^*(dz))$, 
where 
$$
f:C\to \C \big/ \Z^2
$$ 
is a branched covering of a standard
torus, play the role of lattice points in this 
moduli space.

The moduli spaces of holomorphic differentials
and, in particular, their volumes are 
important in ergodic theory, for example, 
for the study of  billiards 
in rational polygons \cite{EMZ}. The exact 
evaluation of the sum \eqref{gen1tr}
and the modular properties
of the answer were of great help for 
the asymptotic analysis performed in \cite{EO}. 
The modular transformation exchanges the
$q\to 1$ limit with the $q\to 0$ limit in 
\eqref{gen1tr}, which means relating large 
partitions to small partitions. This 
is an example of a \emph{mirror} 
phenomenon, see for example the discussion 
in \cite{Dij}.

\subsection{Random partitions from localization}\label{sloc}

A constant source of partition sums in geometry 
is equivariant localization \cite{AtBo}. Partitions index
fixed points of the torus action on the
Grassmann varieties (in the Pl\"ucker embedding, 
these are precisely the vectors $v_\lambda$).
They also index fixed points of the torus
action on $\Hilb_n(\C^2)$, the Hilbert 
scheme of $n$ points in the plane $\C^2$, see
e.g.\ \cite{Hai,Na}.

\subsubsection{Hilbert scheme of points in the plane}

By definition, a point in $\Hilb_n(\C^2)$
is an ideal $I\in \C[x,y]$ of 
codimension $n$ as a linear subspace, such as, for example,
the space of polynomials vanishing at $n$ given
distinct points in the plane. The 
torus $(\C^\times)^2$ acts on $\Hilb_n(\C^2)$
by dilating the coordinates $x$ and $y$. 
Its fixed points are the \emph{monomial ideals},
 that is, ideals spanned
by monomials $x^i y^j$. Monomial ideals $I$
are naturally indexed by partitions of $n$,
namely,  
the set 
$\{(i,j),\,  x^i y^j \notin I\}$
is, essentially, a diagram of a 
partition. 

In equivariant localization, the contribution of
an isolated fixed point $I$ appears with a weight which the 
reciprocal of the product of the weights of 
the torus action on the tangent space $T_I$ at $I$.
Let $\epsilon_1$ and $-\epsilon_2$ be the weights
of the torus action on $\C[x,y]$. Then, up to 
a sign, the natural measure on partitions that arises 
 is the following \emph{Jack polynomials}
deformation $\fM_\textup{Jack}$ of the Plancherel measure. 
\begin{figure}[!htbp]
  \begin{center}
    \scalebox{0.5}{\includegraphics{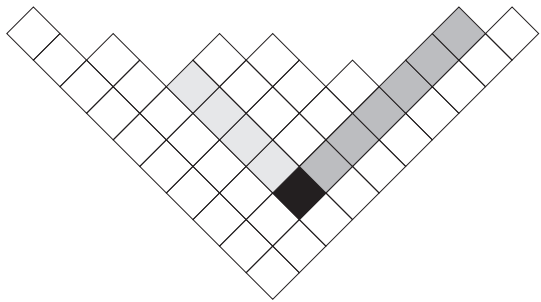}} 
    \caption{The hook, arm, and leg of a square}
    \label{f3}
  \end{center}
\end{figure}

\subsubsection{Jack deformation of the Plancherel measure}

Recall 
the hook-length formula
\begin{equation}
  \label{hookl}
  \frac{\dim \lambda}{|\lambda|!} = 
 \prod_{\square\in \lambda} 
h(\square)^{-1}\,,
\end{equation}
where the product is over all squares $\square$ in the
diagram of $\lambda$ and $h(\square)$ is the 
length of the \emph{hook} of the square $\square$,
see Fig.~\ref{f3}. 

We have $h(\square)=1+a(\square)+l(\square)$, where 
$a(\square)$ and $l(\square)$ are the arm- and
leg-length of $\square$, indicated by 
different shades of gray in Fig.~\ref{f3}. 
In the deformed measure $\fM_\textup{Jack}$ one takes  arm- and
leg-length
with different weights. Concretely, one sets
\begin{equation}
  \label{defMj}
  \fM_\textup{Jack}(\lambda) = \prod_{\square\in \lambda}\frac1
{\Big((1+a(\square))\epsilon_1+l(\square)\epsilon_2\Big)
\Big(a(\square)\epsilon_1+(1+l(\square))\epsilon_2\Big)} \,. 
\end{equation}
Up to an overall factor, $\fM_\textup{Jack}$
clearly depends only on the ratio $\epsilon_1/\epsilon_2$. 
To make $\fM_\textup{Jack}$ a probability measure on 
partitions of $d$, one has to multiply it
by $d !\, (\epsilon_1 \epsilon_2)^d$. 

The measure $\fM_\textup{Jack}$ should be viewed as the 
\emph{general $\beta$} analog of the Plancherel 
measure with $\beta=2\epsilon_1/\epsilon_2$. 
Recall that in random matrix 
theory by ensembles with general $\beta$ one means the
generalization of the measure \eqref{wGUE} 
in which Vandermonde squared is replaced by
$\prod |x_i-x_j|^{\beta}$. Like in 
the random matrix theory, the measure $\fM_\textup{Jack}$
shares some features of Plancherel measure
and lacks others. Most importantly, the
free fermion interpretation is lacking, making,
for example, the computation of the
correlation functions of $\fM_\textup{Jack}$ a difficult
open problem. 

We will see this measure again in Section \ref{sSW}.
Also note that the symmetry $\beta\mapsto 4/\beta$,
of which there are some instances in the random 
matrix theory, is manifest for the discrete 
measure $\fM_\textup{Jack}$.

\section{Schur measure}

\subsection{Definition and correlation functions }

\subsubsection{Schur functions}

A generalization of the Plancherel 
measure, which in many ways resembles 
placing a random matrix into an arbitrary 
potential, is defined as follows. Let
$t_1,t_2,\dots$ be parameters. The 
polynomials
$$
s_\lambda(t) = 
\left(\exp\left(\sum_{n>0} t_n \, \alpha_{-n}
\right) \cdot v_\emptyset, v_\lambda\right)\,,
$$
indexed by partitions $\lambda$, are
 known as the \emph{Schur functions}. Upon 
the substitution
$$
t_k = \frac1k \tr H^k \,,
$$
the polynomial $s_\lambda$ becomes the trace of 
a matrix $H\in GL(n)$ in the irreducible 
representation of $GL(n)$ with highest weight $\lambda$.

As a generalization of \eqref{vPP}, we 
introduce the \emph{Schur measure} on partitions by the
following formula 
\begin{equation}
  \label{defMS}
  \fM_\textup{Schur}(\lambda) =\frac1{\cZ} 
\, s_\lambda(t)\, s_\lambda(\bar t) \,.
\end{equation}
Here $t$ and $\bar t$ are two independent sets of
variables. Choosing one set to be the complex
conjugate of the other is sufficient to 
guarantee that $\fM_\textup{Schur}(\lambda)\ge 0$. This 
positivity, however, is largely irrelevant for what 
follows. 
The normalizing factor $\cZ$ in \eqref{defMS} is given 
by the Cauchy identity
\begin{equation}
  \label{Cauchy}
 \cZ= \sum_\lambda s_\lambda(t)\, s_\lambda(\bar t) = 
\exp\left(\sum_k k\, t_k \bar t_k \right) \,.
\end{equation}
Applying the operator $\sum_{k>0} t_k \partial_{t_k}$ to \eqref{Cauchy}
we see that the expected size $ \lang |\lambda| \rang$ of $\lambda$ with 
respect to the Schur measure equals 
\begin{equation*}
  \lang |\lambda| \rang = \sum_k k^2 t_k \bar t_k \,. 
\end{equation*}

\subsubsection{Schur measure and relative GW theory}

Schur measure naturally arises, for example, in the \emph{relative
Gromov-Witten theory} of $\C^\times=\P^1\setminus\{0,\infty\}$,
see \cite{OP2}. 
This relative Gromov-Witten theory is certain hybrid of 
GW and Hurwitz theory, in which one insist that 
the map $f:C\to \P^1$ has given ramification over
$0,\infty\in \P^1$ (but what precisely having given 
ramifications means when 
$C$ is not smooth is too technical to describe here). 
Informally, these two ramifications can be thought as
in- and out-going states of an interaction described
by the ``worldsheet'' $C$. The variables $t$ and $\bar t$
are the generating function variables coupled
to the cycles of the in- and out-going ramifications. 

\subsubsection{Correlation functions of the Schur measure}

A remarkable property of the Schur measures is that it 
is possible to compute their correlations in closed 
form. Like for random matrices, the $n$-point
correlations are given by $n\times n$ determinants
with a certain correlation kernel $\bK$. Unlike
random matrices, the kernel $\bK$ does not 
involve any delicate objects like polynomials 
orthogonal with an arbitrary weight \cite{D,Me}. On the
contrary, the kernel $\bK$ has a simple 
integral representation in terms of the 
parameters $t$ and $\bar t$, which is 
particularly suited for the steepest descent
asymptotic analysis.  

Introduce the function $T(z)$ by
$$
T(z) = \sum_{k>0} t_n \, z^n -  \sum_{k>0} \bar t_n \, z^{-n} \,.
$$
Let us assume that it converges in some neighborhood of the 
unit circle (the case when $T(z)$ is a polynomial is 
already interesting enough). The correlation function of the Schur measure
were computed in \cite{Oiw} in the following form 
\begin{Theorem} \label{t2}
For any $X\subset \Z+\frac12$ we have
\begin{equation}
  \label{thm2-1}
  \Prob_{\, \fM_\textup{Schur}}\big\{ X\subset \fS(\lambda)\big\}  = \det 
\big[ \bK(x_i,x_j) \big]_{x_i,x_j\in X}\,,
\end{equation}
where the kernel 
$\bK$ is given by 
\begin{equation}
  \label{thm2-2}
  \bK(x,y) = \frac1{(2\pi i)^2} 
\iint_{|w|<|z|} \frac{e^{T(z)-T(w)}}{z-w} \, 
\frac{dz dw}{z^{x+\frac12}{w^{-y+\frac12}}} \,.
\end{equation}
\end{Theorem}

The proof of this formula uses the algebra of the 
infinite wedge representation $\LV$. In the same 
spirit, one shows that the for any fixed set $X$
the sequence correlation functions
$$
\tau^X_n(t,\bar t)=\Prob\big\{ X+n\subset \fS(\lambda)\big\}\,,
$$
where $X+n$ denotes the translation of the set $X$ 
by $n$ lattice spacings, is a sequence of $\tau$-function 
for the Ueno--Takasaki  2-dimensional \emph{Toda hierarchy} with respect
to the two sets of (higher) times $t$ and $\bar t$.

\subsection{Asymptotics and limit shapes}

\subsubsection{Asymptotics of the correlation functions}

Our goal now is to explain how convenient is the 
representation \eqref{thm2-2} for the steepest 
descent analysis. For simplicity, let us assume that 
the variables $\bar t$ are complex conjugate of 
$t$. The interesting asymptotic regime is when 
all variables $t$ grow at the same rate, that is, when
$T(z) = M P(z)$, where $M\to\infty$ and $P(z)$ is 
fixed. As we will see, this implies that the 
typical partition is of length $O(M)$ in both 
directions, and hence contains $O(M^2)$ squares. 
We will investigate $\bK(x_1,x_2)$ in this
limit assuming that 
\begin{equation}
\frac{x_i}{M} \to \bar x\,, \quad  x_1 - x_2 \to \Delta x \,.
\end{equation}
The number $\bar x$
describes our global position on the limit
shape; the number $\Delta x$ is the relative local 
displacement. 

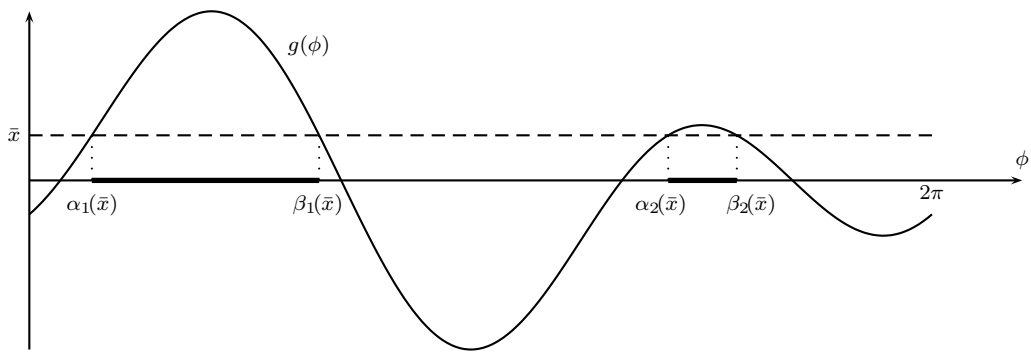
\begin{figure}[htbp]\psset{xunit=1.2cm,yunit=0.75cm}
  \begin{center}
    \begin{pspicture}(0,-3)(11,3)
\scriptsize
\psline{->}(0,-3)(0,3)
\psline[]{->}(0,0)(11,0)
\psline[linestyle=dashed](0,0.8)(10,0.8)
\psline[linewidth=2pt]{-}(.6927911092,0)(3.210934258,0)
\psline[linewidth=2pt]{-}(7.077902513,0) (7.838090754,0)
\psline[linestyle=dotted](.6927911092,0)(.6927911092,0.8)
\psline[linestyle=dotted](3.210934258,0)(3.210934258,0.8)
\psline[linestyle=dotted](7.077902513,0)(7.077902513,0.8)
\psline[linestyle=dotted](7.838090754,0)(7.838090754,0.8)

\rput[t](0.7,-0.25){$\alpha_1\!(\bar x)$}
\rput[t](3.2,-0.25){$\beta_1\!(\bar x)$}
\rput[t](7,-0.25){$\alpha_2\!(\bar x)$}
\rput[t](8,-0.25){$\beta_2\!(\bar x)$}

\rput[r](-0.1,0.8){$\bar x$}
\rput[b](3.1,2.2){$g(\phi)$}
\rput[b](11,0.2){$\phi$}
\rput[t](10,-0.1){$2\pi$}

\psecurve
(-.100000,-.728050)(0.000000,-.601471)(.100000,-.451344)(.200000,-.279460)(.300000,-.088010)(.400000,.120444)(.500000,.343012)(.600000,.576519)(.700000,.817551)(.800000,1.062510)(.900000,1.307672)(1.000000,1.549240)(1.100000,1.783407)(1.200000,2.006418)(1.300000,2.214623)(1.400000,2.404543)(1.500000,2.572919)(1.600000,2.716768)(1.700000,2.833430)(1.800000,2.920614)(1.900000,2.976430)(2.000000,2.999429)(2.100000,2.988621)(2.200000,2.943497)(2.300000,2.864037)(2.400000,2.750718)(2.500000,2.604503)(2.600000,2.426835)(2.700000,2.219613)(2.800000,1.985169)(2.900000,1.726233)(3.000000,1.445895)(3.100000,1.147559)(3.200000,.834893)(3.300000,.511778)(3.400000,.182248)(3.500000,-.149568)(3.600000,-.479507)(3.700000,-.803436)(3.800000,-1.117307)(3.900000,-1.417223)(4.000000,-1.699492)(4.100000,-1.960680)(4.200000,-2.197667)(4.300000,-2.407683)(4.400000,-2.588357)(4.500000,-2.737744)(4.600000,-2.854357)(4.700000,-2.937184)(4.800000,-2.985701)(4.900000,-2.999878)(5.000000,-2.980176)(5.100000,-2.927538)(5.200000,-2.843370)(5.300000,-2.729518)(5.400000,-2.588237)(5.500000,-2.422152)(5.600000,-2.234215)(5.700000,-2.027661)(5.800000,-1.805951)(5.900000,-1.572717)(6.000000,-1.331711)(6.100000,-1.086734)(6.200000,-.841588)(6.300000,-.600006)(6.400000,-.365602)(6.500000,-.141810)(6.600000,.068169)(6.700000,.261415)(6.800000,.435332)(6.900000,.587690)(7.000000,.716658)(7.100000,.820828)(7.200000,.899237)(7.300000,.951379)(7.400000,.977210)(7.500000,.977144)(7.600000,.952047)(7.700000,.903217)(7.800000,.832359)(7.900000,.741560)(8.000000,.633244)(8.100000,.510138)(8.200000,.375217)(8.300000,.231662)(8.400000,.082797)(8.500000,-.067961)(8.600000,-.217166)(8.700000,-.361394)(8.800000,-.497310)(8.900000,-.621717)(9.000000,-.731617)(9.100000,-.824256)(9.200000,-.897178)(9.300000,-.948262)(9.400000,-.975764)(9.500000,-.978343)(9.600000,-.955092)(9.700000,-.905550)(9.800000,-.829715)(9.900000,-.728050)(10.000000,-.601471)(10.100000,-.451344)

\end{pspicture}
    \caption{Construction of the intervals 
$[\alpha_i(\bar x),\beta_i(\bar x)]$}
    \label{fgraphg}
  \end{center}
\end{figure}

The exponentially large term
in the integral \eqref{thm2-2} is $e^{M(S(z)-S(w))}$,
where 
\begin{equation}
  \label{S(z)}
  S(z)=P(z)-\bar x\, \log z \,.
\end{equation}
By our hypothesis, $S(z)$ is purely imaginary on 
the unit circle $|z|=1$ and hence the integrand is
rapidly oscillating there. We want to shift 
the contour of integration in $z$ (resp.\ $w$) 
off the unit circle in the direction of $\mp\grad\Re S$, 
so that to make the integrand exponentially small. 
This direction is given by the sign of 
\begin{equation}\label{zdzS}
g(\phi) - \bar x\,, \quad \textup{where} \quad g(\phi) = \left. 
 z\frac{d}{dz} P(z)\right|_{z=e^{i\phi}}  \,.
\end{equation}
The function $g$ is a real-valued analytic function 
on the circle, and the set of points where $g \ge \bar x$
is a finite union of intervals 
\begin{equation}
  \label{interv}
   \{\phi, g(\phi)\ge \bar x\} = 
\bigsqcup_i
[\alpha_i(\bar x),\beta_i(\bar x)]\,,
\end{equation}
see Fig.~\ref{fgraphg}.
The set \eqref{interv} varies from whole circle
to the empty set 
 as $\bar x$ varies from the minimal to 
the maximal value of the function $g$. These extreme
values mark the edges of the limit shape.

\begin{figure}[!htbp]
  \begin{center}
    \scalebox{0.64}{\includegraphics{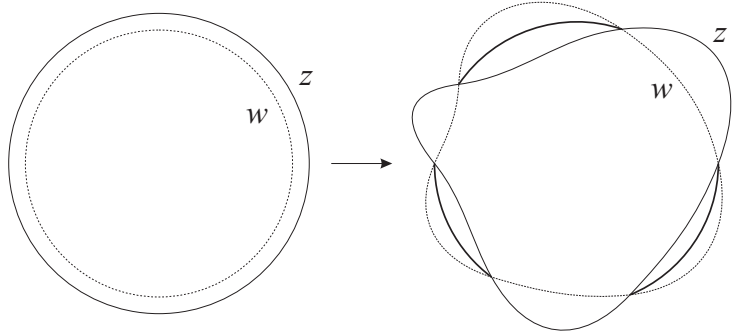}} 
    \caption{Deformation of the integration contours}
    \label{ftwocontours}
  \end{center}
\end{figure}

The integration in \eqref{thm2-2} is along two nested
circles. When we deform the $z$ and $w$ contours 
in the direction of $\mp g(\phi)$ as in Fig.~\ref{ftwocontours},  
we pick up the residue of the integrand at $z=w$ 
whenever we push the $z$-contour inside the the $w$-contour. 
At $z=w$, most of the factors in the integrand cancel out,
so we are left with 
\begin{equation}
  \label{twocont}
 \frac1{(2\pi i)^2} 
 \iint_{|w|<|z|} (\dots)  = 
\frac1{(2\pi i)^2} \iint_\textup{deformed contour}  (\dots)
+ \frac{1}{2 \pi} \sum_i \int_{\alpha_i(\bar x)}^{\beta_i(\bar x)} 
\frac{d\phi}{e^{\phi \Delta x}} \,. 
\end{equation}
By the basics of the steepest descent method, the first
summand in the right-hand side of \eqref{twocont} goes to
$0$ as $M\to\infty$. We obtain 

\begin{Theorem} As $M\to \infty$, we have 
  \begin{equation}
    \label{asbK}
    \bK(x_1,x_2) \to 
\frac{\sum_i \left(e^{-\alpha_i(\bar x)\, \Delta x} 
- e^{-\beta_i(\bar x)\, \Delta x}\right)}
{2\pi \Delta x}\,.
  \end{equation}
\end{Theorem}

Note that \eqref{asbK} is a multi-frequency generalization 
of the discrete sine kernel \eqref{Ksin}. In particular, 
the 1-point function, which determines the density of
particles and, hence, the limit shape, satisfies
\begin{equation}
  \label{dens}
  \Prob\big\{x\in  \fS(\lambda)\big\} = 
K(x,x) \to \frac{\sum_i |\beta_i(\bar x) - \alpha_i(\bar x)|}{2\pi} 
\,.
\end{equation}
This density decreases from $1$ to $0$ as $\bar x$ varies
in $[\min g,\max g]$ and, hence, these
numbers indeed mark the boundary of the limit shape. 

Note that the the density \eqref{dens} is
monotone and hence the  limit shape is always convex. 
Also note that the random process defined by the 
correlation kernel \eqref{asbK}
is translation invariant.

\subsubsection{Example: Plancherel measure}\label{sBDJ}

The Schur measure specializes to the measure $\fM_\textup{PP}$ when
$$
t,\bar t = \left(\sqrt{\xi}, 0,0,\dots\right) \,.
$$
Plugging this into \eqref{thm2-2} leads to the 
\emph{discrete Bessel kernel} defined by 
\begin{align}
  \bK_\textup{Bessel}(x,y;\xi) & = 
\frac1{(2\pi i)^2} 
\iint_{|w|<|z|} \frac{e^{\sqrt{\xi}(z-z^{-1}-w+w^{-1})}}{z-w} \,
\frac{dz\, dw}{z^{x+\frac12}\, w^{-y+\frac12}} \notag \\
&= \sqrt\xi \,\, \frac{J_{x-\frac12} \, J_{y+\frac12} -
J_{x+\frac12} \, J_{y-\frac12}}{x-y} \,,  
  \label{dKB}
\end{align}
where $J_n=J_n(2\sqrt\xi)$ is the Bessel function of
order $n$. This formula is a limit case of the main 
result of \cite{BO2}. It appears as stated in \cite{BOO}
and \cite{J2}. 

We have $M=\sqrt{\xi}$ and $g(\phi)=2\cos(\phi)$. Therefore,
$$
\beta_1(\bar x) = - \alpha_1(\bar x) = \arccos 
\left(\bar x\big/2 \right) \,.
$$ 
Integrating the density $\frac1\pi \arccos 
\frac{\bar x}{2}$, one arrives at the
at the limit shape for the Plancherel measure 
first obtained by Vershik and Kerov \cite{VK1} and Logan and
Shepp \cite{LS}. In the bulk $\bar x\in (-2,2)$ of this 
limit shape, we have the (unscaled) convergence of $\fS(\lambda)$ 
to the \emph{discrete sine-kernel ensemble}
with the correlation kernel 
\begin{equation}
  \label{Ksin}
  \bK_\textup{sin}(x,y;a) = \frac{\sin a(x-y)}{\pi (x-y)} \,,
\quad a= \arccos 
\left(\bar x\big/2 \right)\,.
\end{equation}
In the continuous situation,
the parameter $a$ in \eqref{Ksin} can be scaled away, but on a 
lattice it remains a nontrivial parameter. In fact, 
a finer analysis performed in \cite{BOO} shows that the
we have the same convergence to the discrete
sine kernel for the Plancherel 
measure partitions of fixed size $n$ as $n\to\infty$. 

Near the edge of limit shape, the random process $\fS(\lambda)$
converges, after a suitable scaling
to the Airy ensemble. In our setup, this can be 
seen by analyzing the previously discarded first
summand in the right-hand side of \eqref{twocont}. 
On the edge of the limit shape, $\bar x$ is a critical 
value of $g$ and, hence, the corresponding critical 
point of the action \eqref{S(z)} is degenerate. The Airy kernel 
appears effortlessly in the asymptotics (see, for 
example,  \cite{O3} for a pedagogical exposition). 

The depoissonization analysis performed in \cite{BOO} and 
\cite{J2} shows that the same Airy ensemble asymptotics
remains valid for Plancherel measure on partition
of fixed size $n$ as $n\to\infty$. This
is precisely the statement of a conjecture of Baik, Deift, 
and Johansson, first established in \cite{Orp} by
different means.

\section{Random partitions in Seiberg-Witten theory}\label{sSW}

\subsection{Gauge theory partition function}

\subsubsection{Plancherel measure in a periodic potential}

Fix a period $N$ and let $u_{\frac12},u_{\frac32},\dots,u_{N-\frac12}$ be such
that $\sum u_k = 0$. Consider the period $N$ periodic potential $U$
on the lattice $\Z+\frac12$ defined by $U(x)=u_{x\mo N}$. Define
the  energy
of the particle configuration $\fS(\lambda)$ in the potential $U$ by
\begin{equation}
  \label{defcA}
  U(\lambda) = \sum_{\substack{
   x \in \fS(\lambda),\,\, x > - MN }}
u(x)
 \,.
\end{equation}
This does not depend on the cut-off $M$ as long as $M$ is 
a sufficiently large integer. 

We now define the (poissonized) Plancherel measure
in the periodic potential $U$ by
\begin{equation}
  \label{defMU}
  \fM_U(\lambda) = \xi^{|\lambda|} \, 
\exp\left(\frac1\hbar \, U(\lambda)\right) \, 
\left(\frac{\dim \lambda}{|\lambda|!}\right)^2 \,. 
\end{equation}
The properties of this measure are in many ways parallel to the theory of 
periodically weighted planar dimers, developed in 
\cite{KOS}.

\subsubsection{Instantons and Seiberg-Witten prepotential}

In \eqref{defMU}, we dropped the normalization 
factor $e^{-\xi}$ present in \eqref{defPP} because the 
understanding of the partition function for $\fM_U$ is,
anyway, the main problem in the theory. This is because
this partition function, as shown in Section 5 of \cite{NO},
is essentially the Fourier transform of the $\mathcal{N}=2$
pure supersymmetric $SU(N)$-gauge theory partition functions, as
computed by Nekrasov in \cite{N} via instanton calculus. 

For mathematicians, this gauge theory partition function 
is a generating function for certain integrals over the 
moduli spaces of instantons. These are topologically 
defined finite-dimensional integrals, so  
there is a certain similarity in spirit with Witten's formulation 
of the 2D quantum gravity, discussed in Section \ref{sWitten}. 
The actual geometry of the instanton moduli spaces seems
somewhat more accessible than the geometry of $\Mb_{g,n}$.
The equivariant localization approach to these instanton 
integrals, initiated in \cite{LNS1,LNS2} and completed in \cite{N},
is a far reaching generalization of what we saw in 
Section \ref{sloc}. In particular, partition sums 
appear as sums over fixed points. 

The main expected feature of this partition 
function is that its 
quasiclassical 
$$
\xi\to \infty\,, \quad \hbar\to 0 \,, 
$$
asymptotics 
should be described by the \emph{Seiberg-Witten 
prepotential} \cite{SW1,SW2}. This was indeed demonstrated in \cite{NO}
and this is where the large random 
partitions come in. I refer to \cite{N2} for more on 
the geometrical and physical side of this
computation; here we will focus on purely the random 
partition aspect of it. An introduction to the 
Seiberg-Witten theory for a mathematical audience
can be found in \cite{Don}. It also contains many
further references. A different, non-asymptotic,
approach to the analysis of the partition 
function can be found in \cite{NY}. 

The case of the pure gauge theory is just the 
beginning of the Seiberg-Witten theory. Various
theories with matter lead to related measures
on partitions. They are also considered in \cite{NO}.

\subsection{Asymptotics}

\subsubsection{Quasiclassical scaling and measure concentration}

\begin{figure}[!htbp]
  \begin{center}
    \scalebox{0.7}{\includegraphics{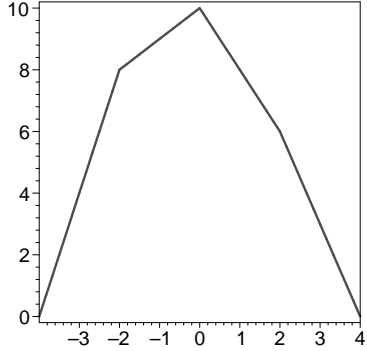}} 
    \caption{The surface tension for $\{u_k\}=\{4,1,-2,-3\}$}
    \label{fsi}
  \end{center}
\end{figure}

Let $\xi$ be very large. Then ${\xi^{|\lambda|}}\big/{|\lambda|!}$
has a sharp peak around $|\lambda|\approx \xi$. For $|\lambda|\approx \xi$,
the weight $\dim\lambda$ can be 
approximately computed as follows. Let $f$
be the profile $f_\lambda(x)$ scaled in both directions by 
$\sqrt{|\lambda|}$, so that to make the area of the scaled 
diagram equal to $1$. By the results of  \cite{LS,VK1,VK3}
\begin{equation}
  \label{logdim}
  \log \,  \xi^{|\lambda|} \,
\left(\frac{\dim \lambda}{|\lambda|!}\right)^2
 \sim - |\lambda|\, E(f) \,,
\end{equation}
where the functional $E(f)$ is defined by 
\begin{equation}
  \label{E}
  E(f)= 2 \iint_{s<t} (1+f'(s))(1-f'(t))\log 2(t-s) \, ds \, dt\,. 
\end{equation}
A direct argument shows that, with the same scaling,
\begin{equation}
  \label{logU}
  \frac1\hbar \, U(\lambda) \sim \frac{\sqrt{|\lambda|}}{2\hbar} \, 
\int \sigma_U(f'(t)) \, dt \,,
\end{equation}
where $\sigma_U(x)$ is a convex continuous
function on $[-1,1]$, which is linear on the 
segments $\big[-1+\frac{2i}{N},-1+\frac{2(i+1)}{N}\big]$,
$i=0,\dots,N-1$, such that the set $\{u_k\}$, sorted
in the decreasing order, is the set of slopes of $\sigma_U$. 
The function $\sigma_U$ has the meaning of \emph{surface
tension}. An example of $\sigma_U$ can be seen in Fig.~\ref{fsi}. 
Note that the surface tension $\sigma_U$ is invariant
under permutations of the $u_i$'s, so in what follows 
we will assume that 
$$
u_{\frac12} > u_{\frac32} > \dots  > u_{N-\frac12} \,.
$$

We see that \eqref{logdim} and \eqref{logU} have the 
same scale when 
$$
\xi \sim \const \, \hbar^{-2} \sim |\lambda| \,. 
$$
In this asymptotic regime, 
the measure $\fM_U$ concentrates on the maximizer
of the functional 
\begin{equation}
  \label{defSf}
  S(f) = - E(f)+ \const \int \sigma_U(f'(t)) \, dt \,.
\end{equation}
This functional is strictly concave and has a 
unique global maximizer $f^\star$. By construction, 
the value $S(f^\star)$
of the functional \eqref{defSf} at its maximizer $f^\star$
dominates the partition function and hence we expect
to identify it with the Seiberg-Witten prepotential. 

Note that the surface tension $\sigma_U(x)$ is not differentiable
at the points $x=-1+\frac{2i}N$. In crystallography, such 
singularities are often called \emph{cusps}. From general 
principles, one expects that the cusps of $\sigma_U(x)$ 
result in \emph{facets} of the maximizer $f^\star$,
that is, the limit shape $f^\star$ develops straight-line pieces
with slopes in the set $\left\{-1+\frac{2i}N\right\}$. 

\subsubsection{Construction of the minimizer}

\begin{figure}[!htbp]
  \begin{center}
    \scalebox{0.7}{\includegraphics{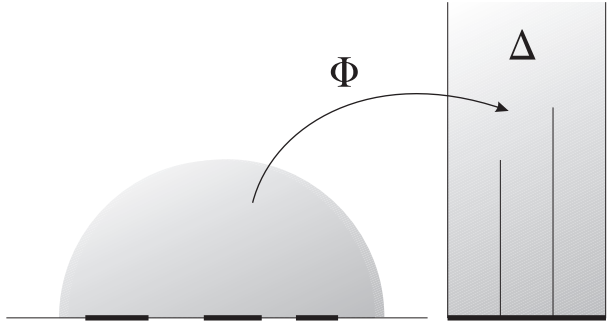}} 
    \caption{The conformal map used in the construction of
the maximizer (here $N=3$)}
    \label{figd}
  \end{center}
\end{figure}

The construction of the maximizer $f^\star$ is 
the following. Let $\Delta$ be the half-strip
$|\Re w|<1$, $\Im w > 0$ in the 
complex plane with vertical slits going 
up from the points $-1+\frac{2i}{N}$, $i=1,\dots,N-1$,
see Fig.~\ref{figd}. The lengths of the slits will
be fixed later. 
Let $\Phi$ be the 
conformal map from the upper half-plane to the 
domain $\Delta$, sending infinity to infinity.
The map $\Phi$ has a certain natural normalization 
at infinity, which fixes it uniquely. 

 Note that $\R$
naturally decomposes into segments of two kinds: those 
(called \emph{bands})
mapped by $\Phi$ to the horizontal parts of $\partial\Delta$,
and those (called \emph{gaps}) mapped to the vertical 
parts of $\partial\Delta$. The slit-lengths of $\Delta$
are found from the $u_i$'s by fixing certain period of the 
differential 
\begin{equation}
  \label{dS}
  dS = z \, d\Phi(z)\,.
\end{equation}
Namely, the integral of $dS$ along the $k$th gap should equal
$u_{k-\frac12}-u_{k+\frac12}$ up to a multiplicative
combination of universal constants such as  $i$ and $\pi$. 
The differential \eqref{dS} will turn out to be 
precisely the \emph{Seiberg-Witten differential}.

We now have the following result from \cite{NO}

\begin{Theorem}\label{tno}
The maximizer $f^\star$ is obtained from the map $\Phi$ by 
the following formula
\begin{equation}
  \label{fphi}
  \frac{d}{dx} f^\star(x)  = \Re \Phi(x+i0) \,,
\end{equation}
where $\Phi(x+i0)$ denotes the natural extension of the map $\Phi$
to the boundary $\R$ of the upper half-plane.
\end{Theorem}

\begin{figure}[!htbp]
  \begin{center}
    \scalebox{0.5}{\includegraphics{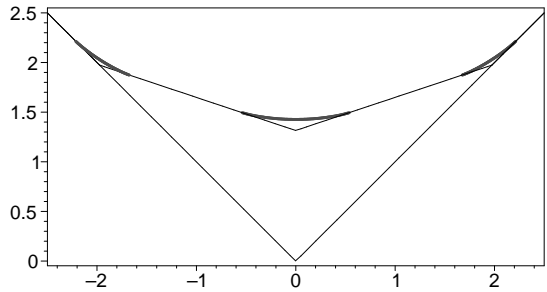}} 
    \caption{An $N=3$ example of the limit shape}
    \label{fielm}
  \end{center}
\end{figure}

 It is clear from \eqref{fphi}
that on the gaps $\frac{d}{dx} f^\star(x)$ is constant, 
which means that the gaps give the facets of the limit
shape. An example of the limit shape is plotted in 
Fig.~\ref{fielm}

\subsubsection{The Seiberg-Witten family of curves}

The conformal map $\Phi$ can be written down 
explicitly, namely it is a linear function of $\log w$, where 
$w=w(z)$ solves the equation
\begin{equation}
  \label{SWC}
  w+ \frac1 w = z^N+0\cdot z^{N-1}+ \dots \,.
\end{equation}
Here dots stand for a certain polynomial of degree $N-2$ in $z$
which is to be found from the slit-lengths and, ultimately, from 
the $u_i$'s. The equation \eqref{SWC} defines an $(N-1)$-dimensional
family of hyperelliptic curves of genus $N-1$. These curves are known
as the \emph{Seiberg-Witten curves\/}. They also arise as
spectral curves in the periodic Toda chain (there is, in fact, a natural 
connection between periodically weighted Plancherel measure 
and periodic Toda chain). 

The $N-1$ gap-periods of the differential \eqref{dS} can be 
taken as local coordinates on the family \eqref{SWC}. The 
main feature of the Seiberg-Witten geometry is that the 
dual band-periods of \eqref{dS} turn out to be the 
dual variables for the Legendre transform of the prepotential
$S(f^\star)$. This follows from Theorem \ref{tno} by a
direct simple computation, see \cite{NO}, and completes the derivation of
the Seiberg-Witten geometry.


\section*{Acknowledgments}

I am grateful to Percy Deift and Michio Jimbo for the 
invitation to Lisbon and to the NSF (grant DMS-0096246)
and Packard foundation for financial support.

\vspace{+10 pt}
\noindent
Princeton University \\
Department of Mathematics\\
Fine Hall, Washington Road, \\
Princeton, New Jersey 08544, U.S.A. \\
okounkov@math.princeton.edu\\

\end{document}